# Coherent X–ray measurements of ion-implantation-induced lattice strains in nano-crystals

Felix Hofmann[1], Edmund Tarleton[2], Ross J. Harder[3], Nicholas W. Phillips[4,5], Jesse N. Clark[6,7], Ian K. Robinson[8,9], Brian Abbey[4], Wenjun Liu[3], Christian E. Beck[2]

[1] Department of Engineering Science, University of Oxford, Parks Road, Oxford, OX1 3PJ, UK

[2] Department of Materials, University of Oxford, Parks Road, Oxford, OX1 3PH, UK

[3] Advanced Photon Source, Argonne National Laboratory, Argonne, Illinois 60439, USA

[4] ARC Centre of Advanced Molecular Imaging, Department of Chemistry and Physics, La Trobe Institute for Molecular Science, La Trobe University, Victoria 3086, Australia

[5] CSIRO Manufacturing Flagship, Parkville 3052, Australia

[6] Stanford PULSE Institute, SLAC National Accelerator Laboratory, Menlo Park, CA 94025, USA

[7] Center for Free-Electron Laser Science, Deutsches Elektronensynchrotron, 22607 Hamburg, Germany

[8] London Centre for Nanotechnology, University College, Gower St, London, WC1E 6BT, UK

[9] Research Complex at Harwell, Rutherford Appleton Laboratory, Didcot, OX11 0FA, UK

**Focussed Ion Beam (FIB) milling is a mainstay of nano-scale machining. By manipulating a tightly focussed beam of energetic ions, often gallium (Ga$^+$), FIB can sculpt nanostructures via localised sputtering. This ability to cut solid matter on the nano-scale has revolutionised sample preparation across the life-, earth- and materials sciences. For example FIB is central to microchip prototyping[1], 3D material analysis[2,3], targeted electron microscopy sample extraction[4,5] and the nanotechnology behind size-dependent material properties[6,7]. Despite its widespread usage, detailed understanding of the functional consequences of FIB-induced structural damage, intrinsic to the technique, remains elusive. Here, we present nano-scale measurements of three-dimensional, FIB-induced lattice strains, probed using Bragg Coherent X-ray Diffraction Imaging (BCDI). We observe that even low gallium ion doses, typical of FIB imaging,**

**cause substantial lattice distortions. At higher doses, extended self-organised defect structures appear, giving rise to stresses far in excess of the bulk yield limit. Combined with detailed numerical calculations, these observations provide fundamental insight into the nature of the damage created and the structural instabilities that lead to a surprisingly inhomogeneous morphology.**

When sufficiently energetic ions collide with a target, they can displace target atoms from their equilibrium lattice positions, causing collision cascades and structural damage[8]. Predicting this ion-implantation damage, its evolution and effect on material properties is not straightforward. For a few systems a semi-quantitative understanding has been achieved[9,10] however in general the effects of ion-implantation remain poorly understood. Yet they have important consequences. For example FIB-milled nano-structures have been used extensively to investigate the size-dependence of material properties, leading to the "smaller is stronger" paradigm[6,7]. However several studies suggest that FIB-induced defects could themselves be the major contributor to this observed scale-dependence of material strength[11,12].

The damage produced by FIB-milling ranges from amorphisation[13] to the generation of lattice defects[12] and formation of intermetallic phases[14]. To examine its effect, detailed measurements of the lattice strains that govern defect interactions are essential. Previously, FIB-induced strains have been inferred by considering the deflection of FIB-milled cantilevers[15]. However, such coarse measurements cannot capture the rich detail of heterogeneous defect distributions that are most important.

Here we report the non-destructive three-dimensional nano-scale characterisation of FIB-milling-induced lattice strains in initially pristine objects. Gold was chosen as a model system since some transmission electron microscopy (TEM) characterisation of ion-implantation-damage is available[16], and near-perfect nano-crystals can be reliably grown[17]. Our experiments use non-destructive Bragg Coherent X-ray Diffraction Imaging (BCDI)[18], where a 3D coherent X-ray diffraction pattern (CXDP), i.e. an oversampled 3D reciprocal space map, is collected from a coherently illuminated single crystal (Figure 1(a)). The CXDP corresponds to the intensity of the 3D Fourier transform of the Bragg electron density of the crystal. By recovering the phase of the CXDP, the real space, complex-valued, Bragg electron density can be reconstructed[18]. Its amplitude provides information about electron density, $\rho(\mathbf{r})$, i.e. the shape of the crystal. Its phase, $\psi(\mathbf{r})$, is linked to displacements, $\mathbf{u}(\mathbf{r})$, of atoms from their ideal lattice positions by $\psi(\mathbf{r}) = \mathbf{q}.\mathbf{u}(\mathbf{r})$, where $\mathbf{q}$ is the scattering vector of the CXDP. By combining at least three crystal reflections with linearly independent $\mathbf{q}$ vectors, $\mathbf{u}(\mathbf{r})$, is recovered[19]. By differentiating $\mathbf{u}(\mathbf{r})$, the lattice strain tensor, $\mathbf{\varepsilon}(\mathbf{r})$, is determined. Thus BCDI allows the 3D nano-scale measurement of both crystal morphology and the full lattice strain tensor.

First we consider the effect of FIB-imaging. Before X-ray measurements nano-crystal A was exposed to a gallium ion dose just sufficient to image the sample (30 keV, 50 pA, 4.2 x $10^4$ ions/µm$^2$). The reconstructed morphology (Figure 1(b)), based on the (1-11) CXDP, is in excellent agreement with scanning electron micrographs (Figure 1(a)). The 3D field of lattice displacements in the [1-11] direction (Figure 1(b) and (c)) shows large displacements near the top, implanted, surface of the crystal. By comparison, the lattice displacements measured in an unimplanted, as-grown crystal are small (Figure 1(d) and (e)), with only slight increases at crystal vertices due to surface energy effects[20]. This suggests that the gallium-ion-bombardment causes the large displacements in crystal A.

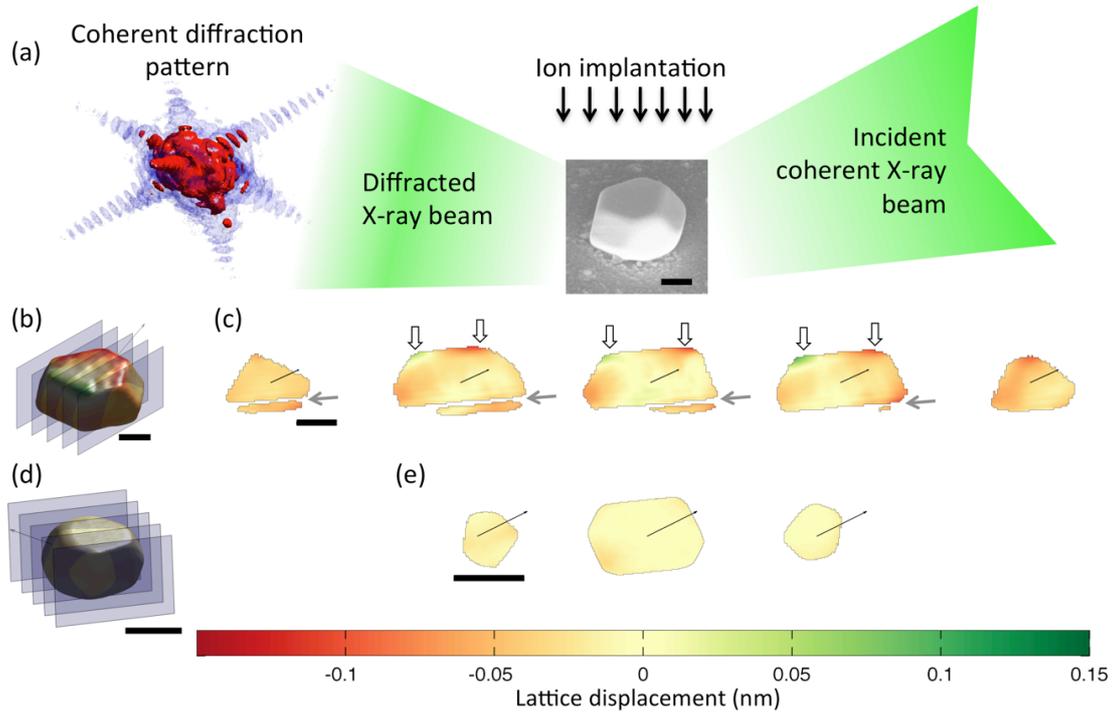

***Figure 1: Lattice displacement fields due to FIB imaging. (a)*** *Schematic of the experimental configuration, SEM micrograph of gold crystal A and 3D rendering of the (1-11) CXDP collected from this crystal. Low dose FIB-imaging (4.2 x $10^4$ gallium ions/$\mu m^2$) was carried out at normal incidence on the top surface of crystal A.* ***(b)*** *Real-space reconstruction of crystal A based on the (1-11) CXDP shown in (a) and coloured by lattice displacement.* ***(c)*** *Cross-sections through the reconstructed displacement field of crystal A (sections shown as blue planes in (b)). Grey arrows show missing intensity due to a twin domain. White arrows show large lattice displacements near the implanted top surface.* ***(d)*** *Real-space reconstruction of a virgin gold nano-crystal colour-coded according to lattice displacement, and reconstructed from a {111} CXDP.* ***(e)*** *Cross-sections through the displacement field of the virgin crystal (sections indicated by blue planes in (d)). Lattice displacements in (b) – (e) are shown in the direction of the scattering vector (thin black arrow). Scale bars are 300 nm in length.*

To further explore these FIB-imaging-induced lattice distortions, CXDPs from five reflections were used to reconstruct the full 3D lattice strain tensor, **ε**(**r**) (supplementary section 1.5). The six independent components of **ε**(**r**) are shown on virtual xy and yz sections through crystal A (Figure 2). $\varepsilon_{yy}(\mathbf{r})$ is large and negative within ~30 nm of the implanted top surface, suggesting a lattice contraction due to gallium-implantation. More subtle strain features are present in the $\varepsilon_{xy}(\mathbf{r})$ (Figure 2(c)) and $\varepsilon_{yz}(\mathbf{r})$ (Figure 2(d)) shear components.

These strains can be understood by comparison with numerical calculations. Using the measured 3D morphology, an anisotropically elastic[21] finite element (FE) model of crystal A was constructed (Figure 2(e), supplementary section 1.6). Simulations using the Stopping Range of Ions in Matter (SRIM) code (supplementary section 1.2)[22] predict a ~20 nm thick damage layer (Extended Data Figure 2). Accordingly a constant volumetric strain, $\varepsilon_v$, is imposed within a 20 nm thick surface layer in the FE model (Extended Data Figure 4). $\varepsilon_v = -3.15 \times 10^{-3}$ provides a good match to the experimentally measured lattice displacement fields. There is striking agreement between calculated strains (Figure 2(f) and (g)) and measured strains (Figure 2(c) and (d)). Not only are the $\varepsilon_{yy}(\mathbf{r})$, $\varepsilon_{xy}(\mathbf{r})$ and $\varepsilon_{yz}(\mathbf{r})$ components well matched, but more subtle features in the other strain components also agree. This demonstrates that even very low gallium ion fluences lead to substantial lattice distortions, and highlights the potential of BCDI for detailed, 3D-resolved nano-scale strain analysis.

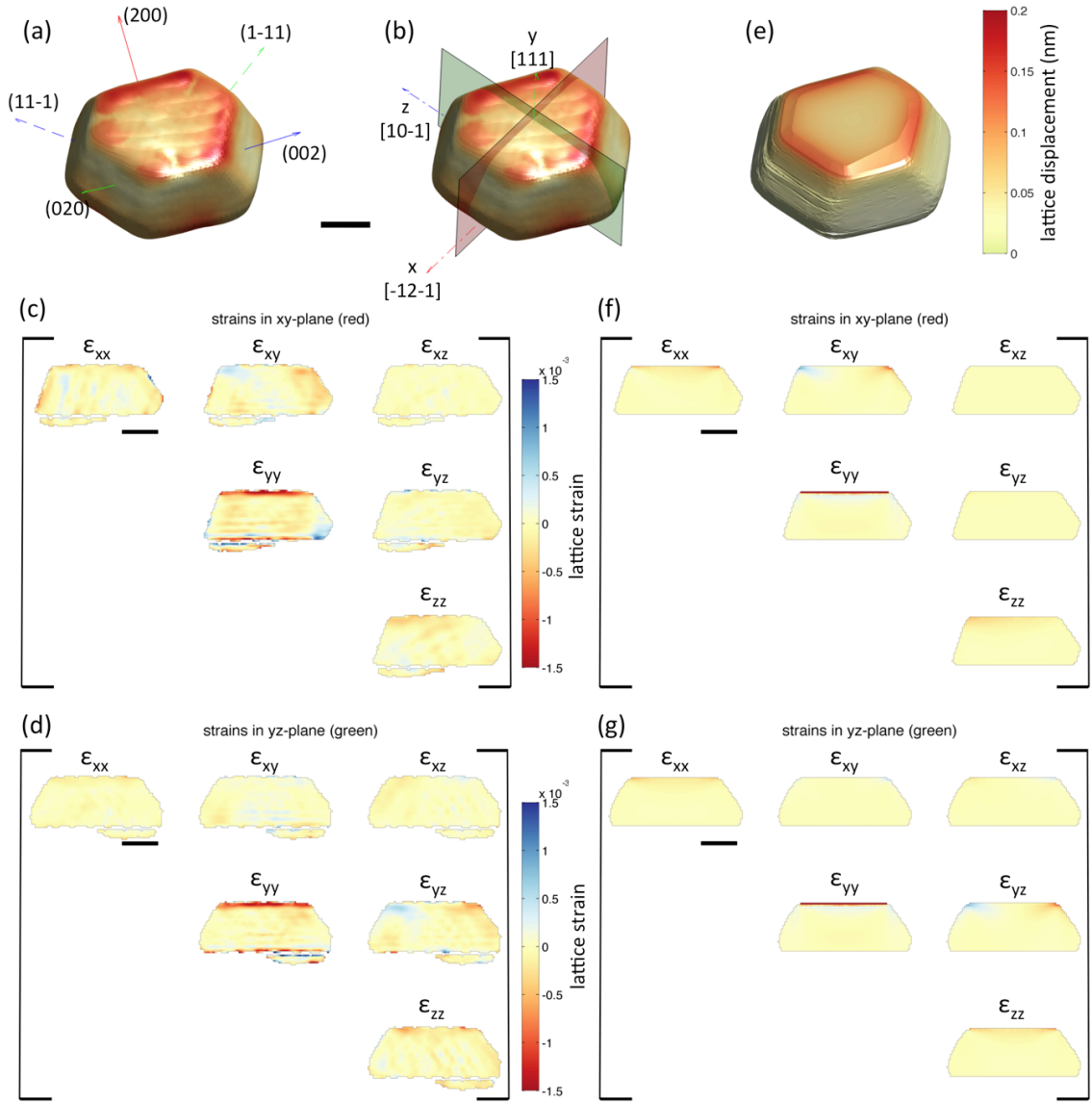

*Figure 2: Full lattice strain tensor in crystal A after low dose gallium-implantation: experiment and simulation. (a) 3D rendering of crystal A coloured by magnitude of the lattice displacement field. Superimposed are the **q** vectors of the five crystal reflections that were measured. (b) Crystal coordinate system used for plotting lattice strains and sections on which strains are plotted. (c) and (d) Maps of the six independent lattice strain tensor components on an xy section (red plane in (b)) and a yz section (green plane in (b)) though crystal A respectively. (e) Finite element model of crystal A showing the predicted displacement field. Extended Data Figure 4 shows the finite element mesh and imposed loading. Calculated lattice strain components are plotted on the same xy (f) and yz (g) planes (shown in (b)) as the experimental data. Scale bars are 300 nm in length.*

It is interesting that gallium-ion-implantation causes a lattice contraction, in contrast to the swelling observed in light-ion-implanted tungsten[10]. Volumetric strain due to defects is given by $\varepsilon_v = \sum_A n(A)\, \Omega_r(A)$, where $n(A)$ and $\Omega_r(A)$ respectively are number density and relaxation volume of defect type $A$[10]. For a gold monovacancy $\Omega_r(V) = -0.27$ [23], while for a self-interstitial atom (SIA) $\Omega_r(SIA) \approx 1.5$ [24]. Hence collision damage in the bulk, involving equal numbers of SIAs and vacancies, will cause a lattice swelling. The observed lattice contraction indicates an excess of vacancies, due to proximity of the crystal surface: vacancies with high migration energy ($\approx$0.75 eV)[25] are retained, while SIAs (migration energy $\approx$ 0.06 eV)[24] escape to the free surface. This agrees with TEM observations of vacancy-cluster-dominated damage in self-ion-implanted gold foils[16]. While TEM is most sensitive to defects >1 nm[26], X-ray diffraction captures the integral effect of all defects via their strain fields. This is important since defects <1 nm can dominate the damage formed[9]. $\varepsilon_v$ and $\Omega_r(V)$ allow a lower bound estimate of ~300 retained vacancies/(gallium ion), while SRIM calculations provide an upper bound of ~400 vacancies/(gallium ion). Thus our measurements allow quantitative assessment of the damage formed, even at very low fluences.

At higher gallium doses a distinctly different behaviour is observed. Nano-crystals B and C were exposed to fluences of 1.3 x 10$^7$ ions/µm$^2$ and 1.5 x 10$^8$ ions/µm$^2$ respectively, causing the removal of ~3 nm and ~40 nm thick surface layers by sputtering, as predicted by SRIM (supplementary section 1.2). Lattice displacements and strains in both crystals were reconstructed using six crystal reflections (Extended Data Figure 5 and Figure 3 respectively). Even for these highly damaged crystals, agreement of the reconstructed morphology and SEM micrographs is excellent (Extended Data Figure 1).

The lattice displacement magnitude of crystal C (Figure 3(a)) shows substantial variations, in contrast to the gradual changes in crystal A (Figure 2(a)). The $\varepsilon_{yy}(\mathbf{r})$ strain (Figure 3(c)) is no longer uniform and negative in the implanted layer, but contains compressive and tensile regions. Similar variations are present in the other strain components.

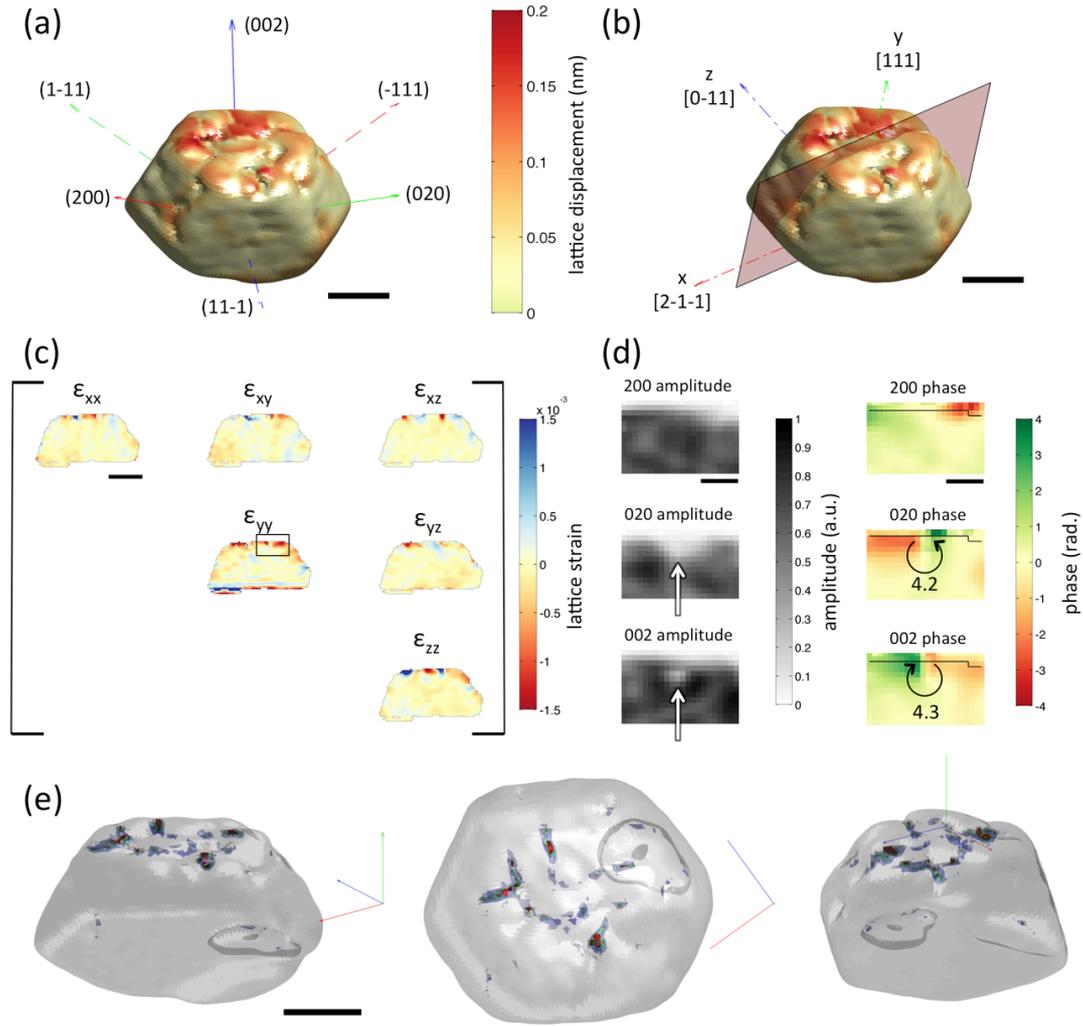

*Figure 3: Displacements, Strains and Stresses in crystal C after FIB milling (gallium fluence 1.5 x 10$^8$ ions/µm$^2$). (a)* 3D rendering of crystal C coloured according to the reconstructed lattice displacement magnitude. Superimposed are the **q** vectors of the six reflections that were measured. *(b)* Crystal coordinates and section plane on which the crystal strains are plotted. *(c)* Maps of the six independent lattice strain tensor components plotted on a xy section through the crystal (red plane in (b)). *(d)* Magnified view of amplitudes and phases of the complex electron density reconstructed from {200} reflections. The region corresponds to that marked by a black box on the $\varepsilon_{yy}$ map in (c) and is centred on a defect. White arrows point to areas of reduced amplitude in the (020) and (002) reflections. Circular arrows indicate the phase jump (in radians) in the (020) and (002) phases. *(e)* Semi-transparent rendering of the outer crystal shape. Superimposed are iso-surfaces of von Mises stress, corresponding to 300 MPa (blue), 400 MPa (green) and 500 MPa (red). Three different viewpoints are shown. Scale bars are 300 nm in length in (a), (b), (c) and (e), and 100 nm in length in (d).

The nature of the underlying crystallographic defects can be explored by considering the amplitudes and phases of the complex electron densities recovered from different crystal reflections. For example the defect in figure 3(d) causes phase jumps of ~4.2 radians in the (020) and (002) reflections, while the (200) reflection shows no phase jump. This suggests a dislocation with burgers vector a/3 [01-1]. Furthermore a local reduction appears in the recovered amplitudes of the (020) and (002) reflections, but not the (200) reflection. This agrees with BCDI observations of "pipes of missing intensity" at dislocation cores[27]. Indeed throughout the crystal, phase features consistent with a/3<110> stair-rod dislocations[28] are visible (supplementary note 2.1). We do not observe any phase features consistent with individual stacking fault tetrahedra, expected to form during ion-bombardment of gold[29]. This is probably because their expected size (5 to 10 nm) is below our spatial resolution.

It is interesting to consider the average strains induced by FIB milling as a function of ion dose. Extended Data Figure 6 shows profiles of $\varepsilon_{xx}$, $\varepsilon_{yy}$, and $\varepsilon_{zz}$ for crystals A, B and C plotted against depth from the implanted surface. $\varepsilon_{yy}$ in the implanted layer of crystal A is approximately twice as large as in crystals B and C, but has much less variation. This suggests that at higher gallium fluences larger defects, as well as the clustering of point defects[10], act to relieve implantation-induced strains by localising lattice distortion.

The ordering of larger defects in crystal C can be visualised by computing the von Mises stress (supplementary section 1.5). Figure 3(e) shows von Mises stresses >500 MPa in the implanted surface layer, greatly exceeding the yield strength of bulk gold (55–200 MPa)[30]. The arrangement of defects in lines differs from TEM observations of uniformly distributed (<100 nm) dislocation loops in FIB-milled copper[12]. The spacing between defect lines (~200 nm) is close to that predicted by the solution, $u_y(\mathbf{r}) = A\sin(kx)\exp(-ky)$, $u_x(\mathbf{r}) = u_z(\mathbf{r}) = 0$, of the biharmonic equation $\Delta^2 \mathbf{u}(\mathbf{r}) = 0$ that governs the deformation of solids. This solution, admissible only at free surfaces, links the length-scale of distortions in the depth-direction (FIB-damage depth) to that of distortions in the lateral direction (see supplementary note 2.2). It is remarkable that this simple deformation model captures the defect self-organisation we observe.

The natural next step is to consider even more extensively machined samples, such as crystal D (supplementary note 2.3) into which a central hole was FIB-milled. The measured strain fields are remarkably complex (Supplementary Data Figure 8) and highlight exciting challenges for the future development of theoretical models able to correctly predict this behaviour.

In summary, BCDI allows unprecedented 3D nano-scale measurements of the full lattice strain tensor even in highly damaged materials, as we demonstrate on FIB-milled gold nano-crystals. Our results show that even low gallium-ion doses, typical of FIB imaging, introduce substantial lattice distortions. At the higher fluences used for FIB-milling, extended self-organised defect structures form that are surprisingly heterogeneous. Combining these measurements with numerical calculations gains us detailed insight into the rich damage microstructure produced by FIB-milling. Our study thus provides a new pathway to understanding the complex nano-structural changes brought about by ion-bombardment.

Methods:

**Sample manufacture:**

Gold crystals were prepared by dewetting a 20 nm thick gold layer, thermally evaporated onto a silicon (Si) substrate with a 2 nm titanium (Ti) adhesion layer (supplementary section 1.1). The resulting crystals range from ≈100 nm to a few µm in size and show facets corresponding to {111} and {100} crystal planes (Figure 1 (a)). No FIB-milling was carried out in the vicinity of the unimplanted reference crystal. FIB-milling of crystals A, B and C was carried out at normal incidence, using a 30 keV, 50 pA gallium ion beam and fluences of $4.2 \times 10^4$ ions/ µm$^2$, $1.3 \times 10^7$ ions/ µm$^2$ and $1.5 \times 10^8$ ions/ µm$^2$ respectively. Crystal D was exposed to a fluence of $4.2 \times 10^4$ ions/ µm$^2$ and a central, nominally 200 nm diameter, region to a fluence of $2.5 \times 10^9$ ions/ µm$^2$. To allow reliable measurement of multiple reflections from crystals A, B, C and D, FIB was used to remove any other gold crystals within a 20 µm radius. Scanning electron micrographs of crystals A, B, C and D are shown in Extended Data Figure 1.

**Ion implantation calculations:**

Ion implantation calculations used the "monolayer collision – surface sputtering" model in the Stopping Range of Ions in Matter (SRIM) code[22]. For the gold target displacement energy of 44 eV, binding energy of 3 eV and surface energy of 3.8 eV were used (supplementary section 1.2). Gallium ions were implanted at normal incidence with an energy of 30 keV, gathering statistics over $10^5$ ions. Each ion was estimated to cause on average ~ 430 target displacements, of which ~30 were replacement collisions. The calculated sputtering rate was ~15.5 gold atoms per gallium ion. For crystal A the amount of material removed by sputtering was negligible. For crystals B and C, an estimated layer of thickness ~3 nm and ~40 nm respectively was removed. Custom MATLAB scripts were used to capture the receding surface effect due to sputtering. The calculated displacement damage and gallium concentration profiles for crystals A, B, C and D, plotted as a function of depth, are shown in Extended Data Figure 2.

**Experimental measurements:**

Synchrotron X-ray micro-beam Laue diffraction at beamline 34-ID-E at the Advanced Photon Source, Argonne National Lab, USA was used to determine the lattice orientations of gold crystals. This served to pre-align crystals for coherent X-ray diffraction measurements at beamline 34-ID-C at the Advance Photon Source. Measurements on the unimplanted reference crystal used an X-ray energy of 9.25 keV, while diffraction patterns from crystals A, B, C and D were collected at 10.2 keV. The X-ray beam size was 1.4 x 2.1 $\mu m^2$ (h x v). For each crystal CXDPs from the following reflections were collected: unimplanted reference: {111}; crystal A: (1-11), (11-1), (200), (020), (002); crystal B: (-111), (1-11), (11-1), (200), (020), (002); crystal C: (-111), (1-11), (11-1), (200), (020), (002); and crystal D: (-111), (1-11), (11-1), (200), (020), (002). Unfortunately the (-111) reflection of crystal A was physically inaccessible. Further details of the experimental measurements are provided in supplementary section 1.3.

**Phase retrieval:**

The phase retrieval algorithm to recover the real-space complex electron density is adapted from published work[27] and is described in more detail in supplementary section 1.4. Briefly, each 3D CXDP pattern was treated independently, using a guided phase retrieval approach with 20 random starts and 5 generations. For each generation 330 phase retrieval iterations were performed using Error Reduction and Hybrid-Input-Output algorithms. Trials using larger numbers of iterations showed no significant further evolution of the solution. Partial coherence effects were accounted for[17], and the normalised mutual coherence functions, recovered for all reflections, are consistent with an almost fully coherent illumination. After the fifth generation a sharpness metric was used to select the three best estimates, which were then averaged to return the reconstructed complex electron density. Finally all reconstructions were transformed into an orthogonal laboratory reference frame with isotropic real-space pixel spacing. Agreement between the reconstructed crystal morphologies and scanning electron micrographs is excellent (Extended Data Figure 1).

**3D reconstruction of lattice displacements, strains and stresses:**

To recover the 3D lattice displacement field, **u(r)**, of a given crystal, any phase wraps in the complex electron densities reconstructed from multiple crystal reflections were unwrapped. Next all reconstructions were rotated into the same sample coordinate frame. The phase of the electron density reconstructed from a particular *hkl* peak, $\psi_{hkl}(\mathbf{r})$, is linked to the scattering vector $\mathbf{q}_{hkl}$ and lattice displacement **u(r)** by $\psi_{hkl}(\mathbf{r}) = \mathbf{q}_{hkl} \cdot \mathbf{u}(\mathbf{r})$. Thus each reconstruction provides a projection of **u(r)** along the corresponding $\mathbf{q}_{hkl}$. If 3 reflections with linearly independent $\mathbf{q}_{hkl}$ are measured, **u(r)** can be reconstructed. Here 5 (crystal A) or 6 (crystals B, C and D) reflections with non-collinear **q** vectors were measured from each crystal. Thus the system of equations is over determined, and a least square fit was used to calculate **u(r)**. The symmetric Cauchy strain tensor, **ε(r)**, is found by differentiating **u(r)**. The strain uncertainty of our measurements, estimated from line profiles of **ε(r)** extracted from crystal A (Extended Data Figure 6), is ~$10^{-4}$. Stresses were computed from **ε(r)** using anisotropic elastic constants for gold[21]. Further details are provided in supplementary section 1.5.

**Finite element simulations:**

Finite element simulations were performed in Abaqus 6.14, using the experimentally determined crystal morphology as a template for generation of the finite element mesh. The custom matlab and python scripts developed for this purpose are available upon request. A global seed size of 10 nm was used, based on mesh dependency studies that showed negligible improvements for finer mesh sizes. The resulting model for crystals A and D are shown in Extended Data Figure 4. Material properties were captured using anisotropic linear elastic constants for gold[21]. A uniform lattice contraction, $\varepsilon_v$, was imposed within a 20 nm thick layer at ion-implanted faces (top face for crystal A, top face and hole wall for crystal D) to represent the effect of ion-implantation damage. $\varepsilon_v$ = -3.15 x $10^{-3}$ provides a good match to the experimentally measured lattice displacement fields in crystal A. Displacements on the bottom surface of the crystals were fixed to capture the substrate effect. A more detailed description of the simulation procedures is provided in supplementary section 1.6.

Acknowledgements:

FH acknowledges funding from the John Fell fund (122/643) and the Royal Society (RG130308). ET acknowledges funding through an EPSRC Early Career Fellowship (EP/N007239/1). JNC acknowledges financial support from the Volkswagen Foundation. Diffraction experiments used the Advanced Photon Source, a U.S. Department of Energy (DOE) Office of Science User Facility operated for the DOE Office of Science by Argonne National Laboratory under Contract No. DE-AC02-06CH11357. This work was supported by the Australian Research Council Centre of Excellence in Advanced Molecular Imaging (CE140100011) www.imagingcoe.org. We also acknowledge funding from the United Kingdom Engineering and Physical Sciences Research Council via programme grants EP/H018921/1 and EP/I022562/1.

Author contributions:

FH designed the project. JNC, IKR and CEB made the samples. FH, RJH, NWP and WL carried out the experiments. FH carried out experimental analysis with input from RJH and JNC. ET carried out the numerical modelling. FH and ET wrote the paper. RJH, NWP, JNC, IKR, BA, WL and CEB read and commented on the manuscript.

Competing financial interest:

The authors declare no competing financial interests.

# Coherent X-ray measurements of ion-implantation-induced lattice strains in nano-crystals: Supplementary Information

Felix Hofmann *[1], Edmund Tarleton[2], Ross J. Harder[3], Nicholas W. Phillips[4,5], Jesse N. Clark[6,7], Ian K. Robinson[8,9], Brian Abbey[4], Wenjun Liu[3], and Christian E. Beck[2]

[1]Department of Engineering Science, University of Oxford, Parks Road, Oxford, OX1 3PJ, UK
[2]Department of Materials, University of Oxford, Parks Road, Oxford, OX1 3PH, UK
[3]Advanced Photon Source, Argonne National Laboratory, Argonne, Illinois 60439, USA
[4]ARC Centre of Advanced Molecular Imaging, Department of Chemistry and Physics, La Trobe Institute for Molecular Science, La Trobe University, Victoria 3086, Australia
[5]CSIRO Manufacturing Flagship, Parkville 3052, Australia
[6]Stanford PULSE Institute, SLAC National Accelerator Laboratory, Menlo Park, CA 94025, USA
[7]Center for Free-Electron Laser Science, Deutsches Elektronensynchrotron, 22607 Hamburg, Germany
[8]London Centre for Nanotechnology, University College, Gower St, London, WC1E 6BT, UK
[9]Research Complex at Harwell, Rutherford Appleton Laboratory, Didcot, OX11 0FA, UK

## 1 Supplementary Methods

### 1.1 Sample manufacture

Gold nano-crystals were manufactured by depositing a 2 nm thick layer of titanium followed by a 20 nm thick layer of gold on a silicon wafer using thermal evaporation. The sample was then annealed at 1273 K in air for 10 h, following which the gold film had dewetted and formed nanocrystals.

Focussed ion beam milling (FIB) was used to clear a circular area of 40 $\mu$m outer diameter around crystals A,B,C and D. This is necessary for reliable identification of specific crystals and to allow the unambiguous measurement of several Bragg reflections from the same crystal. FIB work was carried out on a Zeiss Auriga dual beam microscope. Spatial alignment of the electron and ion-beam was carried out at the wafer edge, far from the gold crystals under study. Using the electron beam suitable crystals approximately 1 $\mu$m in diameter were identified. Initial FIB imaging and milling used a 30 kV, 50 pA Ga ion beam. A single FIB image of a 30 x 30 $\mu$m$^2$ area, centered on the crystal of interest, was recorded with a fluence of $6.8 \times 10^{-15}$ C/$\mu$m$^2$, corresponding to $4.2 \times 10^4$ ions/$\mu$m$^2$. This was the minimum fluence for which an image of sufficient quality to identify the crystal of interest could be achieved. Based on this alignment image an annular milling scan with 2.5 $\mu$m inner and 10 $\mu$m outer diameter and a fluence of $5 \times 10^{-10}$ C/$\mu$m$^2$ was used to remove other gold crystals in the immediate vicinity of the crystal of interest. Next the following additional FIB exposures of crystals A, B, C and D were carried out:

- Crystal A: No additional FIB exposure. Total dose (due to FIB imaging): **$4.2 \times 10^4$ ions/$\mu$m$^2$**

- Crystal B: Additional FIB exposure of whole crystal to fluence of $2 \times 10^{-12}$ C/$\mu$m$^2$ ($1.3 \times 10^7$ ions/$\mu$m$^2$). Total dose: **$1.3 \times 10^7$ ions/$\mu$m$^2$**

- Crystal C: Additional FIB exposure of whole crystal to fluence of $2.4 \times 10^{-11}$ C/$\mu$m$^2$ ($1.5 \times 10^8$ ions/$\mu$m$^2$). Total dose: **$1.5 \times 10^8$ ions/$\mu$m$^2$**

---

*felix.hofmann@eng.ox.ac.uk

- Crystal D: Central, nominally 200 nm diameter region exposed to fluence of $4 \times 10^{-10}$ C/$\mu$m$^2$ ($2.5 \times 10^9$ ions/$\mu$m$^2$). Total dose for the edge of crystal D: $\mathbf{4.2 \times 10^4}$ **ions/$\mu$m$^2$**. Total dose for the hole in crystal D: $\mathbf{2.5 \times 10^9}$ **ions/$\mu$m$^2$**ions)

Finally, using a 30 kV, 1 nA Ga ion beam, an annular region of 8 $\mu$m inner and 40 $\mu$m outer diameter was exposed to a fluence of $5 \times 10^{-10}$ C/$\mu$m$^2$ to remove any other gold crystal within a 20 $\mu$m radius of the crystal of interest. High resolution scanning electron micrographs of crystals A, B, C and D after FIB milling are shown in Extended Data Figure 1.

## 1.2 Ion implantation simulations

Ion implantation calculations were carried out using the Stopping Range of Ions in Matter (SRIM) code [1]. For the gold displacement energy 44 eV, binding energy 3 eV, surface energy 3.8 eV were used [2]. Implantation of gallium ions was modeled with 30 keV ions at normal incidence, gathering statistics over $10^5$ implanted ions and using the "monolayer collision - surface sputtering" model.

The predicted sputtering rate of 15.5 gold atoms per injected gallium ion was used to estimate the material removed by different gallium ion fluences. For crystal A material removal due to sputtering is negligible. For crystals B and C we estimate removal of a $\sim$ 3 nm and $\sim$ 40 nm thick surface layer respectively. Each gallium ion is also predicted to cause $\sim$ 430 target displacements, of which $\sim$ 30 are replacement collisions, meaning that $\sim$ 400 vacancies are generated per implanted ion. Using the SRIM-calculated damage and implanted-ion distributions, the damage and injected ion profiles for the different gallium fluences were calculated. Importantly these calculations, carried out in MATLAB, account for the receding-surface effect due to sputtering. The results are shown in Extended Data Figure 2.

The calculated profiles for crystals C and crystal D (hole) are very similar. In both cases a layer in excess of the gallium-ion-penetration depth has been removed by sputtering, meaning that both implantation and damage profiles are expected to be independent of fluence.

## 1.3 Coherent diffraction experiments

White beam Laue diffraction measurements at beamline 34-ID-E at the Advanced Photon Source, Argonne National Lab, USA, were used to determine the lattice orientations of crystals A, B, C and D prior to coherent diffraction measurements. A detailed description of the 34-ID-E instrument is provided elsewhere [3, 4]. Using a monochromatic X-ray beam, focused to a size of 0.6 $\times$ 0.7 $\mu$m$^2$ (h x v) on the sample by KB mirrors, fluorescence measurements were used to identify the spatial position of each crystal. Then, switching over to a polychromatic beam (5 - 30 keV), Laue diffraction patterns from each crystal were collected and fitted using the LaueGo software (J.Z. Tischler, tischler@aps.anl.gov) to determine the lattice orientation, captured by the **UB** matrix [5].

Coherent diffraction measurements were carried out on beamline 34-ID-C at the Advanced Photon Source, Argonne National Lab, USA. A schematic of the angular degrees of freedom at this beamline is shown in Extended Data Figure 3. The incident, monochromatic X-ray beam was focussed to a size of 1.4 x 2.1 $\mu$m$^2$ (h x v) on the sample using KB mirrors. An in situ confocal microscope was used to position crystals A, B, C and D in the X-ray beam. The angular positions required to place specific crystal reflections in diffraction condition were determined based on the **UB** matrix from Laue diffraction measurements. The actual diffraction peaks were reliably found within less than 1$°$ of the calculated positions. The reflections that were measured from each crystal are listed in the second column of Extended Data Table 1.

Diffraction patterns were recorded on a Medipix2 area detector with a 256 x 256 pixel matrix and a pixel size of 55 $\mu$m. For crystals A, B, C and D the detector was positioned at a distance of 1.85 m from the sample and an X-ray energy of 10.2 keV ($\lambda$ = 0.121 nm) was used. 3D coherent X-ray diffraction patterns (CXDP) were recorded by rotating the crystal, covering an angular range of 0.4$°$ in $\theta$ and recording an image every 0.0025$°$ with an exposure time of 1 s. For the unimplanted reference crystal (Fig. 1(d) and (e) in main text) the sample-to-detector distance was 0.635 m and an X-ray energy of 9.25 keV ($\lambda$ = 0.134 nm) was used. CXDPs were recorded by rotating through an angular range of 1.5$°$ in $\theta$ in steps of 0.01$°$ with an exposure time of 0.5 s. To optimise the signal to noise level of the CXDPs, multiple repeated scans of each reflection were measured. Summing over repeated shorter scans is preferable to simply collecting one scan with a longer exposure as it allows correction for sample drift between each scan. The number of times each crystal reflection was measured is listed in column three of Extended Data Table 1.

Using a 3D version of the algorithm described by Guizar-Sicairos et al. [6] the multiple scans taken of each crystal reflection were aligned such as to maximise their cross-correlation coefficient. Scans with a cross-correlation coefficient greater than 0.99 were summed up to produce the CXDP of a specific reflection. Column four in Extended Data Table 1 lists the number of scans that were included in this sum for each reflection. A 3D rendering of the (1-11) CXDP from crystal A is shown in Fig. 1(a) in the main text.

## 1.4 Phase retrieval

The phase retrieval code is adapted from published work [7]. The CXDP of each crystal reflection was treated independently using a guided phase retrieval approach [8] with 20 random starts and 5 generations. The best estimate selection was based on a sharpness metric, previously shown to yield the most truthful reconstruction for strained samples [7].

A low-to-high resolution phasing scheme was used, phasing low spatial resolution data in the first generation, which is then used to seed reconstructions of progressively higher resolution data in later generations [7]. Artificially low spatial resolution data was generated by multiplying the 3D CXDPs with a 3D gaussian of width $\sigma$, given as a fraction of the total array size. $\sigma = 0.1$ and $\sigma = 0.55$ were used for generations 1 and 2 respectively. From generation 3 onward full resolution data was used.

For each generation 330 phase retrieval iterations were performed consisting of a pattern of 10 iterations of Error Reduction (ER) and 40 iterations Hybrid Input-Output (HIO) [9] repeated six times, followed by a final 30 iterations of ER. Reconstructions with larger numbers of iterations per generation showed no visible evolution of the result. The object returned at the end of each generation was the average of the final 10 iterations for each start. At the end of the last generation the returned image was the average of the 3 best estimates (from an initial population of 20). The support was updated every 5 iterations using the shrinkwrap algorithm [10].

Partial coherence of the illumination was accounted for using a previously established method for BCDI measurements [11] based on the iterative Richardson-Lucy algorithm [12]. The normalized mutual coherence functions, recovered for all reflections, are consistent with an almost fully coherent illumination.

Spatial resolution of the reconstructions was determined by taking the derivative of line profiles of the crystal-air-interface and fitting these with a Gaussian. For each reconstruction six profiles (2 in each spatial direction) were measured. The lowest resolution value for each reconstruction is recorded in column five of Extended Data Table 1.

After reconstruction, phase ramps, which correspond to uniform lattice contraction or expansion and are not of interest here, were removed by re-centering the Fourier transform of the complex electron density. Finally reconstructions were transformed into an orthogonal laboratory frame with the z-axis aligned with the X-ray beam direction, the y-axis pointing vertically up and the x-axis direction following the right hand rule (see Extended Data Figure 3). In this reference frame reconstructions were returned with an isotropic real-space pixel spacing of 14.51 nm for crystals A, B, C and D, and 4.16 nm for the unimplanted reference crystal.

## 1.5 Reconstruction of 3D stress and strain fields

The 3D strain fields of crystals A, B, C and D were reconstructed in MATLAB based on the complex electron density recovered from multiple reflections of each crystal. Phase wraps in any of the reconstructions were unwrapped using the algorithm developed by Cusack et al. [13], propagating outwards from a reference position, here chosen as the centre of mass of each crystal.

The scattering vector, $\mathbf{q}_{hkl}$, of a measured reflection $hkl$, is given by $\mathbf{q}_{hkl} = \mathbf{s}_{hkl} - \mathbf{s}_0$. Here $\mathbf{s}_0 = \hat{\mathbf{s}}_0(2\pi/\lambda)$ and $\mathbf{s}_{hkl} = \hat{\mathbf{s}}_{hkl}(2\pi/\lambda)$, where $\hat{\mathbf{s}}_0$ is a unit vector along the incident beam direction. $\hat{\mathbf{s}}_{hkl}$ is a unit vector from the sample position to the detector centre, and is given by $\hat{\mathbf{s}}_{hkl} = \mathbf{R}_y(\delta_{hkl})\mathbf{R}_x(\gamma_{hkl})\mathbf{s}_0$, where $\delta_{hkl}$ and $\gamma_{hkl}$ respectively are the detector angles associated with a specific $hkl$ reflection. $\mathbf{R}_x$, $\mathbf{R}_y$ and $\mathbf{R}_z$ are rotation matrices about the x, y, and z axes and follow a right handed convention.

The angular position of the sample during diffraction measurements is captured by the rotation angles of the sample stack, $\phi_{hkl}$, $\chi_{hkl}$ and $\theta_{hkl}$. This rotation is captured by a rotation matrix $\mathbf{R}_{hkl} = \mathbf{R}_y(\theta_{hkl})\mathbf{R}_z(\chi_{hkl})\mathbf{R}_x(\phi_{hkl})$. By pre-multiplying the coordinates of the different $hkl$ reflections of the same crystal by $\mathbf{R}_{hkl}^T$, all reconstructions were rotated into the same sample coordinate frame. The first column of Extended Data Figure 1 shows a superposition of crystal morphologies reconstructed from the different reflections measured for each crystal (see Extended Data Table 1). Agreement of the morphologies is excellent, and the average morphology of each crystal (column two of Extended Data Figure 1) very closely matches scanning electron micrographs of the crystals (column three of Extended Data Figure 1).

The phase, $\psi_{hkl}(\mathbf{r})$, of the complex electron density reconstructed from each reflection is given by $\psi_{hkl}(\mathbf{r}) = \mathbf{u}(\mathbf{r}) \cdot \mathbf{q}_{hkl}$, where $\mathbf{u}(\mathbf{r})$ is the lattice displacement field. $\mathbf{u}(\mathbf{r})$ in each crystal was recovered by minimising:

$$E(\mathbf{r}) = \sum_{hkl} \left( \mathbf{u}(\mathbf{r}) \cdot \mathbf{q}_{hkl} - \tilde{\psi}_{hkl}(\mathbf{r}) \right)^2, \tag{S1}$$

where $\tilde{\psi}_{hkl}$ is the phase of the complex electron density reconstructed from experiments and the sum was carried out over all $hkl$ reflections from a given crystal.

3D lattice strain fields, $\varepsilon(\mathbf{r})$, in crystals A, B, C and D were computed from $\mathbf{u}(\mathbf{r})$ by numerical differentiation and using the Cauchy strain tensor format:

$$\varepsilon(\mathbf{r}) = \begin{bmatrix} \frac{\partial u_x}{\partial x} & \frac{1}{2}\left(\frac{\partial u_x}{\partial y} + \frac{\partial u_y}{\partial x}\right) & \frac{1}{2}\left(\frac{\partial u_x}{\partial z} + \frac{\partial u_z}{\partial x}\right) \\ \frac{1}{2}\left(\frac{\partial u_y}{\partial x} + \frac{\partial u_x}{\partial y}\right) & \frac{\partial u_y}{\partial y} & \frac{1}{2}\left(\frac{\partial u_y}{\partial z} + \frac{\partial u_z}{\partial y}\right) \\ \frac{1}{2}\left(\frac{\partial u_z}{\partial x} + \frac{\partial u_x}{\partial z}\right) & \frac{1}{2}\left(\frac{\partial u_z}{\partial y} + \frac{\partial u_y}{\partial z}\right) & \frac{\partial u_z}{\partial z} \end{bmatrix}, \quad (S2)$$

where x, y and z refer to the crystal coordinates as shown in main-text Figs. 2(b) and 3(b) for crystals A and C, and in Suppl. Figs. 5(b) and 8(b) for crystals B and D respectively. Plots of strains in these figures always show the upper triangle of the symmetric Cauchy strain tensor.

For all four crystals line profile of strain as a function of distance from the implanted surface were extracted from the 3D strain datasets. The results for crystals A, B and C are plotted in Extended Data Figure 6. Those for the top surface and the central milled hole in crystal D are shown in Extended Data Figure 9. For the central hole in crystal D strains in cylindrical polar coordinates are plotted. $\varepsilon_{rr}$ corresponds to the surface normal component, while $\varepsilon_{yy}$ and $\varepsilon_{\theta\theta}$ are the in-plane strain components. Interestingly Extended Data Figure 9 shows significantly different implantation-induced lattice strains for surfaces implanted with ions at normal incidence (crystal D top surface Extended Data Figure 9(c)) and at glancing incidence (crystal D hole surface Extended Data Figure 9(d)).

The 3D stress fields, $\sigma(\mathbf{r})$, were computed by rewriting $\varepsilon(\mathbf{r})$ in Voigt notation and then pre-multiplying by the gold stiffness tensor rotated into the x, y and z crystal reference frame (see Eqn. S7).

Von Mises stress, $\sigma_{vM}(\mathbf{r})$, was calculated from $\sigma(\mathbf{r})$ using the standard expression [14]:

$$\sigma_{vM}(\mathbf{r}) = \sqrt{\frac{3}{2}\sigma'(\mathbf{r}) : \sigma'(\mathbf{r})}. \quad (S3)$$

Here $\sigma'(\mathbf{r})$ is the deviatoric stress tensor, given by $\sigma'(\mathbf{r}) = \sigma(\mathbf{r}) - \frac{1}{3}\left(\text{tr}\sigma(\mathbf{r})\right)\mathbf{I}$, and $\mathbf{I}$ is the identity matrix.

## 1.6 Finite Element Model

The finite element mesh was generated by importing the cloud point data of the reconstructed crystal morphology (see second column of Extended Data Figure 1) into MATLAB and rotating the coordinate system from the sample frame to crystal coordinates. This ensures the top free surface normal [111] is along the $y$ axis (see main text Fig. 2(b) and Extended Data Figure 8(b) respectively for crystals A and D). The advantage is that the top surface of the crystal is approximately flat and lies in the $x - z$ plane. The geometry was reproduced by extracting a series of 30 slices in the $x - z$ plane, equally spaced along the $y$ direction ($\Delta y = 14.51$ nm) and mapping a convex polygon to the crystal contour in each slice. Then every polygon was simplified to have at most 30 vertices. Deciding which vertices to keep was based on their salience, defined by the length of and angle between the two edges connected to them, $L_1 L_2 \theta_{12}$. An Abaqus python script was developed to read in the polygon coordinates from a text file and generate a series of wire parts which were joined using lofting to generate a 3D part. A 20 nm layer was then partitioned to represent the gallium-implanted layer on the top free surface, and, in the case of crystal D, the free surface of the FIB-milled hole. A global seed size of 10 nm was used to generate the finite element mesh. This seed size was found to be sufficient for convergence and resulted in 230528 nodes and 1274427 linear tetrahedral C3D4 elements for crystal A and 254021 nodes and 1390649 elements for crystal D. 3D visualisations of both models in Extended Data Figure 4 show the finite element mesh and the layer partitioned off to represent the ion-implanted material.

Anisotropic elasticity was used as the anisotropy parameter, $A = 2c_{44}/(c_{11} - c_{12})$, of gold is 2.85 [15]. For crystal A the model $\mathbf{x}$, $\mathbf{y}$, $\mathbf{z}$ coordinate system, is defined in terms of lattice coordinates, i.e. basis vectors aligned with the unit cell edges, as $\mathbf{x} = [-1, 2, -1]/\sqrt{6}$, $\mathbf{y} = [1, 1, 1]/\sqrt{3}$ and $\mathbf{z} = [1, 0, -1]/\sqrt{2}$ (see also Fig. 2(b) in the main text). The rotation matrix $\mathbf{R}$ from lattice coordinates to the model coordinate frame is then given by:

$$\mathbf{R} = \begin{bmatrix} \mathbf{x} \\ \mathbf{y} \\ \mathbf{z} \end{bmatrix} = \begin{bmatrix} -0.4082 & 0.8165 & -0.4082 \\ 0.5774 & 0.5774 & 0.5774 \\ 0.7071 & 0 & -0.7071 \end{bmatrix} \quad (S4)$$

That is to say if $\mathbf{v}$ is a vector in lattice co-ordinates then $\mathbf{v}' = \mathbf{R}\mathbf{v}$ is the same vector expressed in the model coordinate system ($\mathbf{x}$, $\mathbf{y}$, $\mathbf{z}$). The stiffness tensor for a cubic material has 3 unique non-zero components. For gold literature values $c_{11} = 192.9$ GPa, $c_{44} = 41.5$ GPa, $c_{12} = 163.8$ GPa were used [15]. In Voigt notation the stiffness matrix in the lattice coordinate system is

$$\mathbf{D} = \begin{bmatrix} c_{11} & c_{12} & c_{12} & 0 & 0 & 0 \\ c_{12} & c_{11} & c_{12} & 0 & 0 & 0 \\ c_{12} & c_{12} & c_{11} & 0 & 0 & 0 \\ 0 & 0 & 0 & c_{44} & 0 & 0 \\ 0 & 0 & 0 & 0 & c_{44} & 0 \\ 0 & 0 & 0 & 0 & 0 & c_{44} \end{bmatrix} \tag{S5}$$

which needs to be expressed in the model coordinate system. The stress tensor in the crystal system $\boldsymbol{\sigma}$ is expressed in the model system as $\boldsymbol{\sigma}' = \mathbf{R}\boldsymbol{\sigma}\mathbf{R}^T$ which in Voigt notation is written as $\boldsymbol{\sigma}' = \mathbf{T}\boldsymbol{\sigma}$ where $\boldsymbol{\sigma} = [\sigma_{11}, \sigma_{22}, \sigma_{33}, \sigma_{12}, \sigma_{13}, \sigma_{23}]^T$ and the transformation matrix, $\mathbf{T}$, is

$$\mathbf{T} = \begin{bmatrix} R_{11}^2 & R_{12}^2 & R_{13}^2 & 2R_{11}R_{12} & 2R_{11}R_{13} & 2R_{12}R_{13} \\ R_{21}^2 & R_{22}^2 & R_{23}^2 & 2R_{21}R_{22} & 2R_{21}R_{23} & 2R_{22}R_{23} \\ R_{31}^2 & R_{32}^2 & R_{33}^2 & 2R_{31}R_{32} & 2R_{31}R_{33} & 2R_{32}R_{33} \\ R_{11}R_{21} & R_{12}R_{22} & R_{13}R_{23} & R_{11}R_{22} + R_{12}R_{21} & R_{11}R_{23} + R_{13}R_{21} & R_{12}R_{23} + R_{13}R_{22} \\ R_{11}R_{31} & R_{12}R_{32} & R_{13}R_{33} & R_{11}R_{32} + R_{12}R_{31} & R_{11}R_{33} + R_{13}R_{31} & R_{12}R_{33} + R_{13}R_{32} \\ R_{21}R_{31} & R_{22}R_{32} & R_{23}R_{33} & R_{21}R_{32} + R_{22}R_{31} & R_{21}R_{33} + R_{23}R_{31} & R_{22}R_{33} + R_{23}R_{32} \end{bmatrix} \tag{S6}$$

where $R_{ij}$ are the components of $\mathbf{R}$. The stiffness matrix $\mathbf{D}$ can also be transformed using $\mathbf{T}$ as $\mathbf{D}' = \mathbf{T}\mathbf{D}\mathbf{T}^T$ which can be obtained by considering the invariance of the elastic energy which is proportional to $\boldsymbol{\sigma}^T\boldsymbol{\varepsilon} = \boldsymbol{\sigma}'^T\boldsymbol{\varepsilon}' = \boldsymbol{\sigma}^T\mathbf{T}^T\boldsymbol{\varepsilon}'$ therefore the strain vector in Voigt notation must transform as $\boldsymbol{\varepsilon}' = \mathbf{T}^{-T}\boldsymbol{\varepsilon}$. Finally $\boldsymbol{\sigma}' = \mathbf{D}'\boldsymbol{\varepsilon}' = \mathbf{D}'\mathbf{T}^{-T}\boldsymbol{\varepsilon} = \mathbf{T}\mathbf{D}\boldsymbol{\varepsilon}$. Therefore $\mathbf{D}'\mathbf{T}^{-T} = \mathbf{T}\mathbf{D}$ hence

$$\mathbf{D}' = \mathbf{T}\mathbf{D}\mathbf{T}^T = \begin{bmatrix} 219.9 & 145.8 & 154.8 & -12.7 & 0 & 0 \\ 145.8 & 228.8 & 145.8 & 0 & 0 & 0 \\ 154.8 & 145.8 & 219.8 & 12.7 & 0 & 0 \\ -12.7 & 0 & 12.7 & 23.5 & 0 & 0 \\ 0 & 0 & 0 & 0 & 32.5 & 12.7 \\ 0 & 0 & 0 & 0 & 12.7 & 23.5 \end{bmatrix} \text{GPa} \tag{S7}$$

is the required stiffness matrix for crystal A expressed in the model $(\mathbf{x}, \mathbf{y}, \mathbf{z})$ coordinate system. The rotation matrix $\mathbf{R}$ for crystal D differs from that for crystal A only by an additional rotation about the $y$ axis of $2\pi/3$ which produces a different $\mathbf{T}$ but the same stiffness matrix $\mathbf{D}'$ as in Eqn. S7.

The boundary conditions for both models were $\mathbf{u} = 0$ on the bottom surface. The mechanical loading produced by the ion-implantation damage was represented by applying an isotropic volumetric contraction of $\varepsilon_v = -3.15 \times 10^{-3}$ in the 20 nm surface layer. The value of $\varepsilon_v$ was determined by matching the displacement magnitude measured on the top surface of crystal A. The same $\varepsilon_v$ was used for crystal D and the strain maps were then compared along two of the principle planes (see main text Fig. 2 and Extended Data Figure 8 respectively for crystals A and D). To allow accurate comparison of the fields predicted by the model with experiments, all visualisations were computed in Matlab using the same colour scale for both experimental data and simulations.

## 2 Supplementary Notes

### 2.1 Analysis of defects in crystal C

The defects that cause the large strain variations observed in the implanted layer of crystal C can be explored in more detail by considering the amplitude and phase of the complex electron density recovered from different crystal reflections. Extended Data Figure 7 shows a 3D rendering of crystal C with the $\mathbf{q}_{hkl}$ vector directions of the six reflections that were measured superimposed. Amplitude and phase of the complex electron density reconstructed from each reflection are shown on two slices though the crystal. In slice I two defects, (a) and (b) can be identified. Defect (a), also described in the main text (see Fig. 3(d)), manifests itself as a phase jump in the (1-11), (11-1), (020) and (002) reflections. The magnitude of the phase jump varies between 4.1 and 4.3 radians, as shown in Extended Data Figure 2 and listed in Extended Data Table 2. No phase jump from defect (a) is visible in the (-111) and (200) reflections. The magnitude of the expected phase jump, $\Delta\psi_{hkl}$, when considering a burger's circuit around a dislocation with burgers vector $\mathbf{b}$ is:

$$\Delta\psi_{hkl} = \mathbf{b} \cdot \mathbf{q}_{hkl}. \tag{S8}$$

The $\mathbf{q}_{hkl}$ for the six measured reflections, with respect to the undistorted crystal, are:

- (-111) reflection: $\mathbf{q}_{[-111]} = \frac{2\pi}{a}[\text{-}111]$
- (1-11) reflection: $\mathbf{q}_{[1-11]} = \frac{2\pi}{a}[1\text{-}11]$
- (11-1) reflection: $\mathbf{q}_{[11-1]} = \frac{2\pi}{a}[11\text{-}1]$
- (200) reflection: $\mathbf{q}_{[200]} = \frac{2\pi}{a}[200]$
- (020) reflection: $\mathbf{q}_{[020]} = \frac{2\pi}{a}[020]$
- (002) reflection: $\mathbf{q}_{[002]} = \frac{2\pi}{a}[002]$

This suggests that defect (a) is a dislocation with $\mathbf{b} = \frac{a}{3}[01\text{-}1]$. Similar analysis of the other defects shows that (c) and (d) are also dislocations with $\mathbf{b} = \frac{a}{3}$ <110> (see Extended Data Table 2). These so-called stair-rod dislocations are formed through the interaction of two Shockley partial dislocations [16]. For example $\frac{a}{6}[21\text{-}1]+\frac{a}{6}[\text{-}21\text{-}1]\to \frac{a}{3}[01\text{-}1]$ which is energetically favourable and sessile. In supplementary movie 1 several further phase features consistent with stair-rod dislocations can be seen. We also note that the amplitude maps show a local reduction in the reflections where phase jumps are observed. For example defect (c) causes phase jumps in the (-111), (1-11), (200), and (020) reflections. Amplitude reductions, at the location of defect (c), are seen in these reflections, while no phase signature or reduced amplitude are observed for the (11-1) and (002) reflections. This agrees with previous BCDI measurements where dislocations were found to be associated with "pipes of missing intensity" [7].

For defect (b) the analysis is less conclusive, with inconsistent phase jumps observed in the different reflections. This is most likely due to close proximity of defect (b) to a further defect, seen as a reduction in (200) amplitude in close proximity to the dip in amplitude from defect (b) (see slice I in Extended Data Figure 7).

A further question concerns the visibility of stacking fault tetrahedra (SFTs). In gold SFTs are expected to form in large numbers during ion irradiation. Their anticipated size of 5 - 10 nm [17] is substantially less than our spatial resolution. The six SFT edges are associated with stair-rod dislocations with $\mathbf{b} = \frac{a}{6}$ <110>, and the burgers vector of each edge is orientated in a different one of the six possible <110> directions [18]. This means that a SFT would produce a phase signature in the complex electron density recovered from any crystal reflection (i.e. for a complete SFT no reflection exists where $\mathbf{b} \cdot \mathbf{q}_{hkl} = 0$ for all SFT edges). Thus we conclude that it is unlikely that the larger defects we observe are associated with SFTs.

## 2.2 Surface solution to the biharmonic equation

Considering compatibility of strains, it can be shown that the displacement field, $\mathbf{u}(\mathbf{r})$, in an elastic body must obey the biharmonic equation [19]:

$$\Delta^2 \mathbf{u}(\mathbf{r}) = 0, \tag{S9}$$

where $\Delta = \frac{\partial^2}{\partial x^2} + \frac{\partial^2}{\partial y^2} + \frac{\partial^2}{\partial z^2}$. A simple solution to Eqn. S9 is of the form:

$$\begin{aligned} u_x(\mathbf{r}) &= u_z(\mathbf{r}) = 0, \\ u_y(\mathbf{r}) &\propto \sin(kx)e^{-ky}, \end{aligned} \tag{S10}$$

where $k$ is a constant. In bulk samples the requirement for displacements to remain finite precludes this solution as $\lim_{y\to-\infty}(\exp{(-ky)}) = \infty$. However in a halfspace with surface in the $xz$ plane and occupying the region $y \geq 0$ the requirement for finite displacements is met. It is interesting to note that in Eqn. S10 $k$ links the length scale of strains in the $y$-direction to the length scale of strains in the $x$-direction.

For crystal C, $k = 0.03$ nm$^{-1}$ provides a good fit to the $\epsilon_{yy}$ profile as a function of depth from the implanted surface ($y$-direction) shown in Extended Data Figure 6(e). This suggests a lengthscale $L = 2\pi/k = 210$ nm associated with strain fields in the plane of the implanted surface. This is in quite good agreement with the spacing between defect lines observed in crystal C (Fig. 3(e) in the main text).

## 2.3 Extensive FIB-machining: Crystal D

To explore more extensive FIB-machining a central hole was FIB-milled into crystal D (see supplementary section 1.1). BCDI measurements of six lattice reflections were used to reconstruct displacements and lattice strains in this crystal (Extended Data Figure 8), with excellent agreement of crystal morphology from BCDI with SEM images (Extended Data Figure 1). The maps of lattice strain (Extended Data Figure 8 (c) and (d)) show that large strains are no longer confined to the vicinity of implanted surface, but appear in the bulk of the nano-crystal. Furthermore there are large strain variations on the surfaces of the milled hole, most visible in the $\varepsilon_{yy}(\mathbf{r})$ strain component, that point to the presence of defects similar to those in crystal C.

These measurements can be compared to predictions from a 3D, anisotropically elastic, finite element model of crystal D (see supplementary section 1.6). To capture the effect of FIB-induced strains, a volumetric lattice strain, $\varepsilon_v$, was imposed within a 20 nm layer at the crystal top surface, and on the inside faces of the hole (Extended Data Figure 4). Using $\varepsilon_v = -3.15 \times 10^{-3}$, based on crystal A, the calculated lattice displacement magnitude (Extended Data Figure 8(e)) agrees reasonably well with that measured experimentally (Extended Data Figure 8(a)). Both show large lattice displacements concentrated near the top (111) crystal surface.

The simulated strain maps (Extended Data Figure 8(f) and (g)) show negative hoop strains ($\varepsilon_{zz}(\mathbf{r})$ in the xy plane, $\varepsilon_{xx}(\mathbf{r})$ in the yz plane) and negative $\varepsilon_{yy}(\mathbf{r})$ strain close to implanted surfaces, in fairly good agreement with experiments. However the predicted large radial strains near the surface of the central hole ($\varepsilon_{xx}(\mathbf{r})$ in the xy plane, $\varepsilon_{zz}(\mathbf{r})$ in the yz plane) are not observed in experiments. The reason appears to be a change in the swelling behaviour between the normal incidence (top of crystal D) and glancing incidence (hole of crystal D) ion implantation cases. This is highlighted by lattice strain profiles extracted from crystal D as a function of distance from the implanted surface (Extended Data Figure 9). The profiles for the top surface of crystal D (Extended Data Figure 9(c)) show that $\varepsilon_{yy}$, the strain component normal to the surface, is large and negative within the implanted layer, just as in crystals A, B and C. However, the strain profiles for the hole in crystal D (Extended Data Figure 9(d)) show a markedly different behaviour: $\varepsilon_{rr}$, the strain component normal to the hole surface, is positive in the implanted layer. The underlying changes in defect microstructure that lead to this behaviour are not clear.

The strain profiles in Extended Data Figure 9 show that ion-implantation-induced strains in crystal D extend substantially further into the bulk than in crystals A, B and C. This behaviour is partly captured by our FE calculations which show substantial stains in the crystal bulk, similar to experiments, but smaller in magnitude (see $\varepsilon_{xx}(\mathbf{r})$, $\varepsilon_{xy}(\mathbf{r})$ and $\varepsilon_{zz}(\mathbf{r})$ in the xy plane and $\varepsilon_{xx}(\mathbf{r})$ and $\varepsilon_{zz}(\mathbf{r})$ in the yz plane). An important role here is played by the presence of larger crystal defects that are not included in our model. Considering the experimental $\varepsilon_{yy}(\mathbf{r})$ strain components (Extended Data Figure 9(c) and (d)), the strain fields associated with these defects can be seen to extend more than 100 nm into the crystal bulk, highlighting that the residual strain state within the crystal is dominated by FIB-milling induced damage.

# Supplementary References

| Crystal | Reflection | Number of measurements | Measurements averaged | Resolution (nm) |
|---|---|---|---|---|
| A | 1-11 | 30 | 18 | 27 |
|   | 11-1 | 30 | 24 | 26 |
|   | 020 | 30 | 26 | 26 |
|   | 200 | 30 | 23 | 26 |
|   | 002 | 30 | 27 | 26 |
| B | 1-11 | 20 | 9 | 24 |
|   | 11-1 | 20 | 14 | 29 |
|   | -111 | 20 | 16 | 25 |
|   | 002 | 20 | 17 | 20 |
|   | 200 | 20 | 11 | 22 |
|   | 020 | 20 | 16 | 19 |
| C | -111 | 30 | 28 | 31 |
|   | 11-1 | 30 | 27 | 29 |
|   | 1-11 | 18 | 14 | 25 |
|   | 020 | 30 | 22 | 19 |
|   | 200 | 30 | 26 | 23 |
|   | 002 | 20 | 18 | 29 |
| D | 11-1 | 20 | 17 | 28 |
|   | -111 | 20 | 12 | 29 |
|   | 1-11 | 20 | 16 | 26 |
|   | 020 | 20 | 12 | 25 |
|   | 200 | 20 | 9 | 29 |
|   | 002 | 20 | 14 | 18 |
| Unimplanted | 111 | 33 | 30 | 6.3 |

Extended Data Table 1: Summary of the experimental measurements. Column 2 lists the reflections measured for each crystal, column 3 the number of times each reflection was measured and column 4 the number of scans included in the averaged CXDP for each reflection. The spatial resolution of the reconstruction of each reflection is reported in column 5.

| Defect | Phase jump in different reflections (rad) | | | | | | Burgers vector |
|---|---|---|---|---|---|---|---|
| | | (-111) | (1-11) | (11-1) | (200) | (020) | (002) | |
| a | measured | 0 | -4.1 | 4.2 | 0 | 4.2 | -4.3 | $\frac{a}{3}$[01-1] |
| | calculated | 0 | $-\frac{4}{3}\pi$ | $\frac{4}{3}\pi$ | 0 | $\frac{4}{3}\pi$ | $-\frac{4}{3}\pi$ | |
| b | measured | -3.4 | 3.4 | 0 | 3.4 | -2 | 0 | |
| | calculated | | | | | | | |
| c | measured | -4.3 | 4.6 | 0 | 4.6 | -4.5 | 0 | $\frac{a}{3}$[1-10] |
| | calculated | $-\frac{4}{3}\pi$ | $\frac{4}{3}\pi$ | 0 | $\frac{4}{3}\pi$ | $-\frac{4}{3}\pi$ | 0 | |
| d | measured | -4.2 | 0 | 4.3 | 4.1 | 0 | -4.4 | $\frac{a}{3}$[10-1] |
| | calculated | $-\frac{4}{3}\pi$ | 0 | $\frac{4}{3}\pi$ | $\frac{4}{3}\pi$ | 0 | $-\frac{4}{3}\pi$ | |

Extended Data Table 2: Summary of four defects analysed in sample C, shown in Extended Data Figure 7. Column 1 shows the defect designator. Columns 2 - 8 show the measured phase jump and calculated phase jump for each defect in all six reflections. The last column list the burgers vector likely to be associated with each defect.

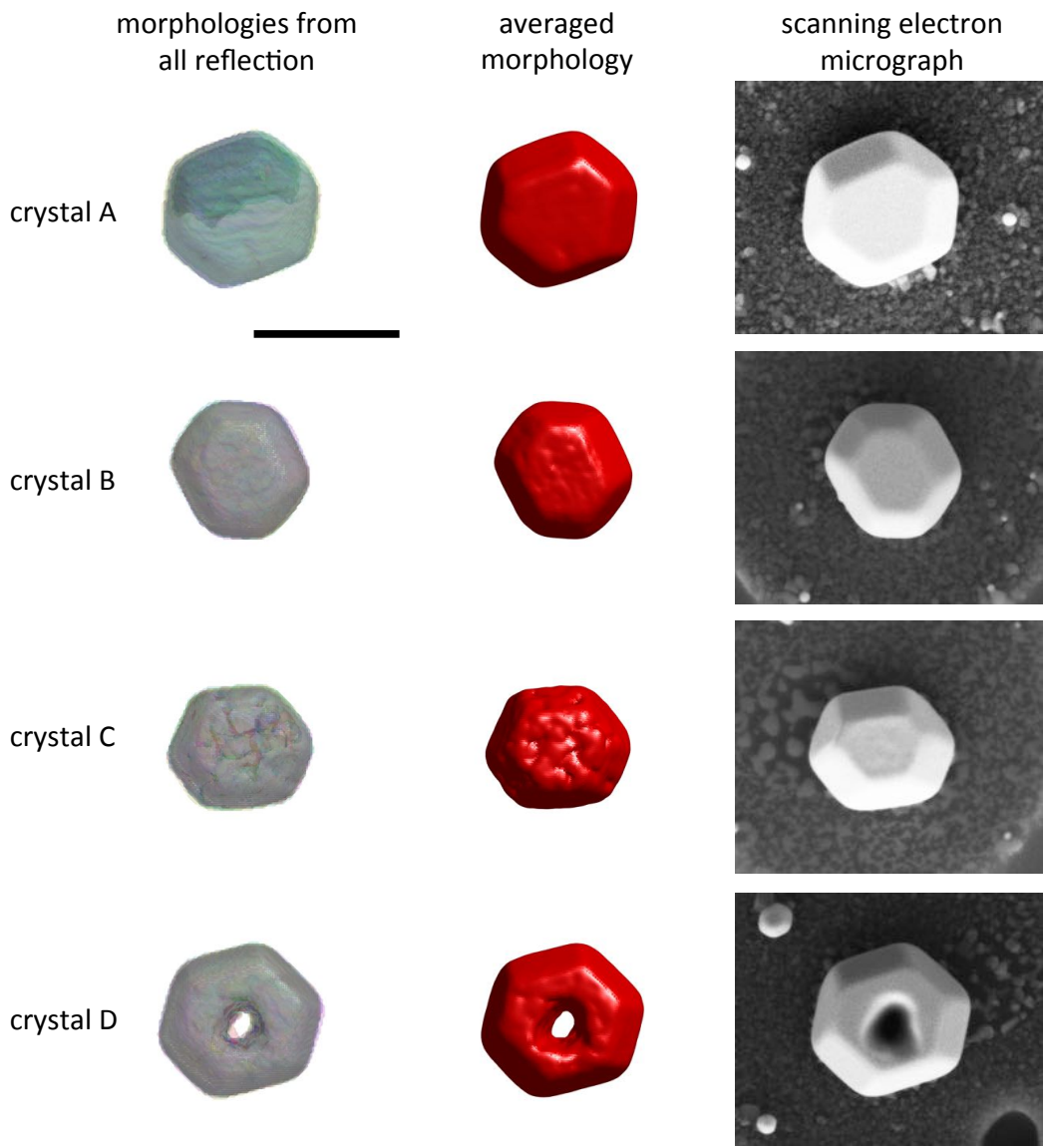

Extended Data Figure 1: Comparison of sample morphologies recovered using BCDI and scanning electron micrographs of crystals A, B, C and D. The first column shows, for each crystal, a superposition of the morphologies recovered from different crystal reflections using BCDI. The second column shows the average morphology of each crystal determined by averaging over the morphologies found from different lattice reflections of that crystal. Scanning electron microscopy images of all four crystals are shown in the third column. All plots are shown at the same magnification and using the same viewpoint. The scalebar corresponds to 1 $\mu$m.

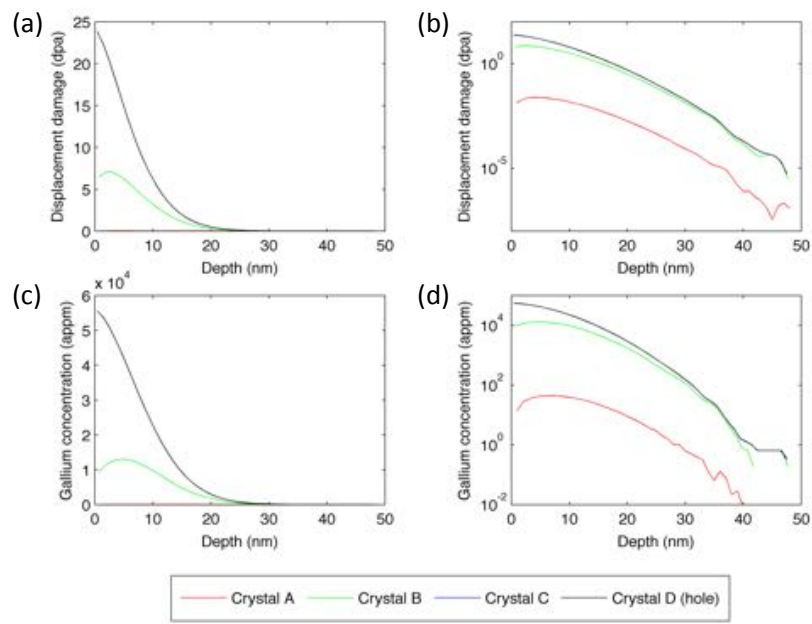

Extended Data Figure 2: Displacement damage and gallium implantation profiles calculated using SRIM. (a) Displacement damage, in displacements per atom (dpa), plotted on a linear scale. (b) Displacement damage plotted on a logarithmic scale. (c) Implanted gallium, in atomic parts per million (appm), plotted on a linear scale. (d) Implanted gallium plotted on a logarithmic scale.

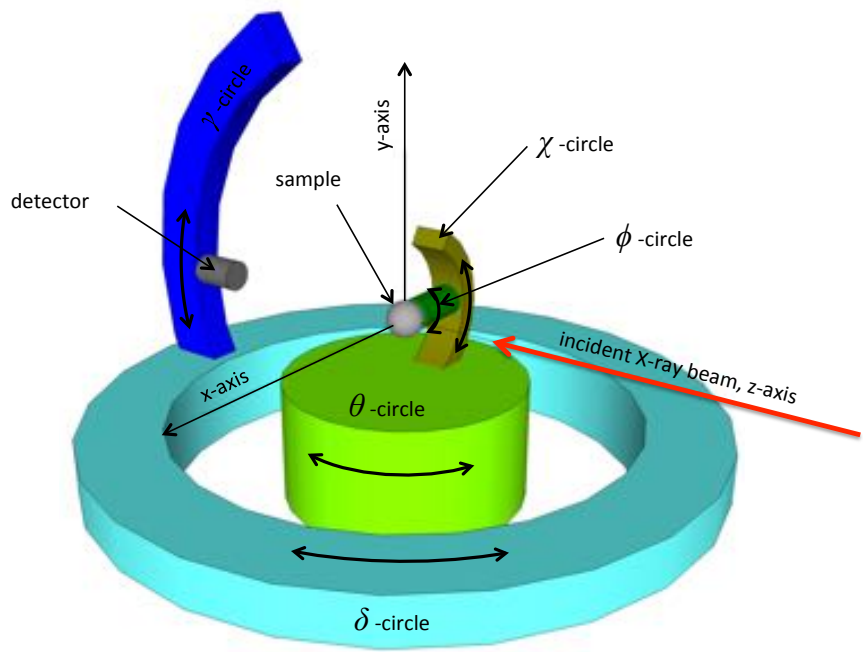

Extended Data Figure 3: Schematic of the coordinate system used at beamline 34-ID-C (Advanced Photon Source, USA) and the rotational degrees of freedom available for sample and detector alignment. All rotation directions follow the right hand rule.

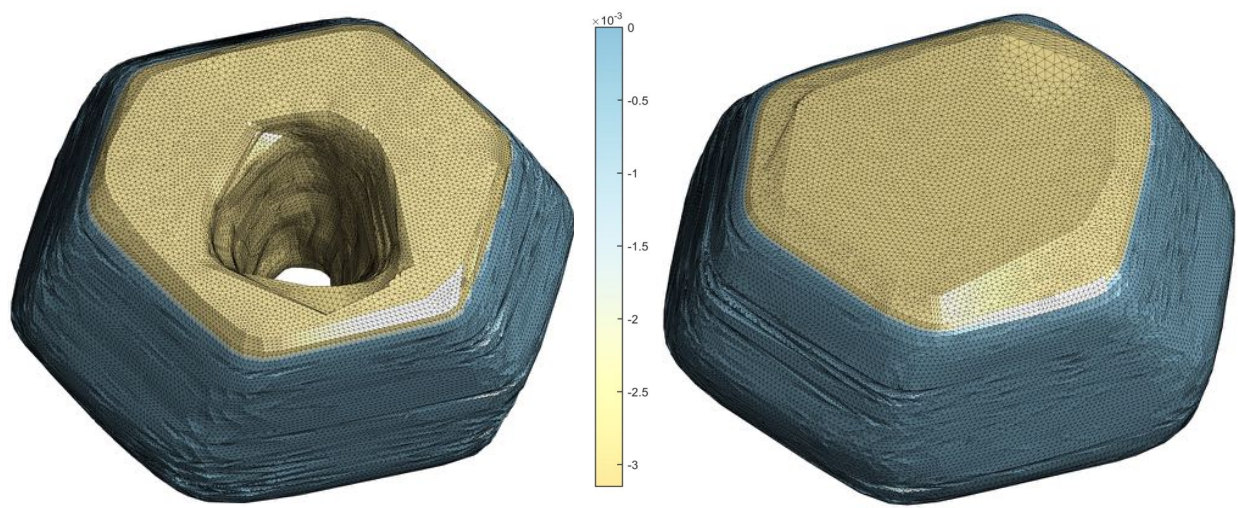

Extended Data Figure 4: 3D visualisation of crystal D and crystal A finite element models. The model mesh is shown and the models are coloured according to the imposed volumetric strain loading, $\varepsilon_v$, showing the lattice contraction applied to the implanted surface layer.

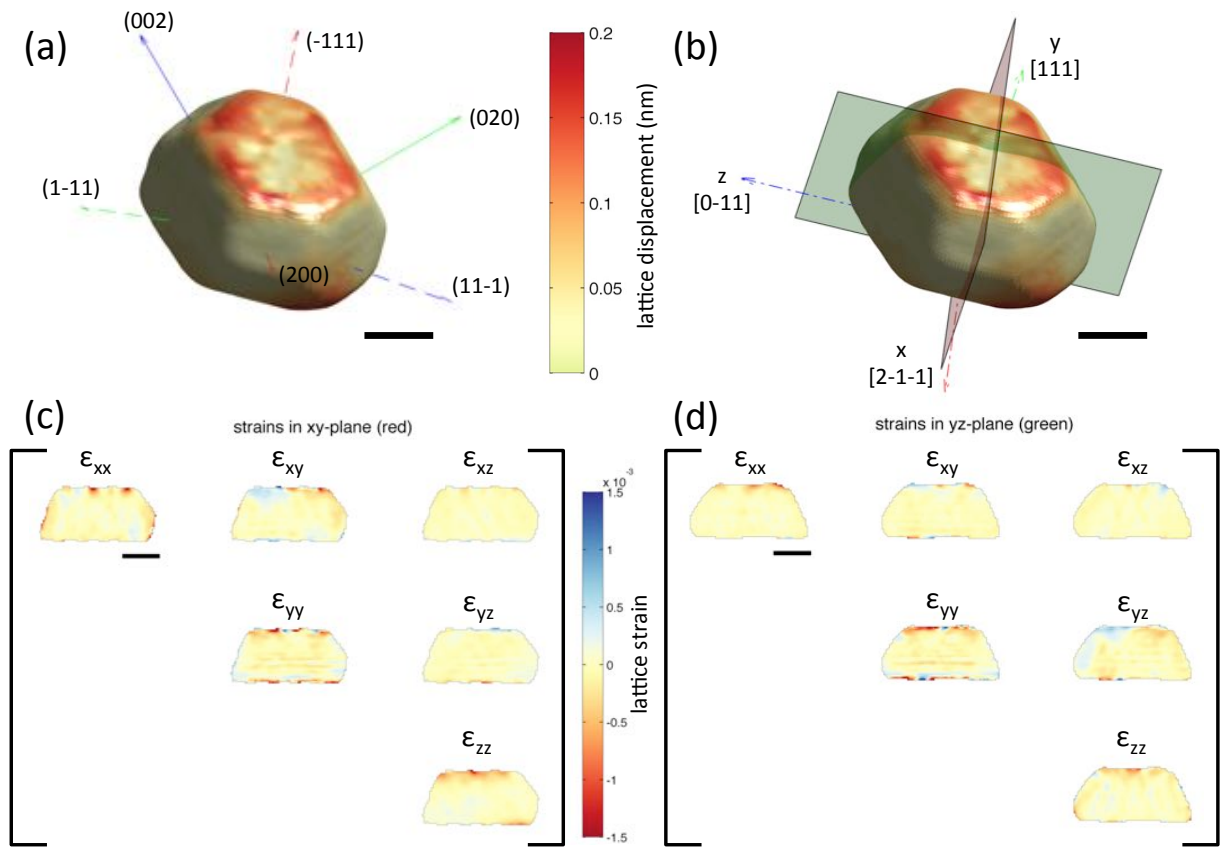

Extended Data Figure 5: Full lattice strain tensor in crystal B after Ga-ion implantation. (a) 3D rendering of crystal B coloured according to the magnitude of the lattice displacement field. Superimposed are the q vectors of the 6 crystal reflections that were measured. (b) Crystal coordinate system used for plotting of lattice strains. (c) Maps of the six independent lattice strain tensor components on an xy section through crystal B (red plane in (b)). (d) Maps of 6 strain tensor components on a yz section though crystal B (green plane in (b)). Scale bars are 300 nm in length.

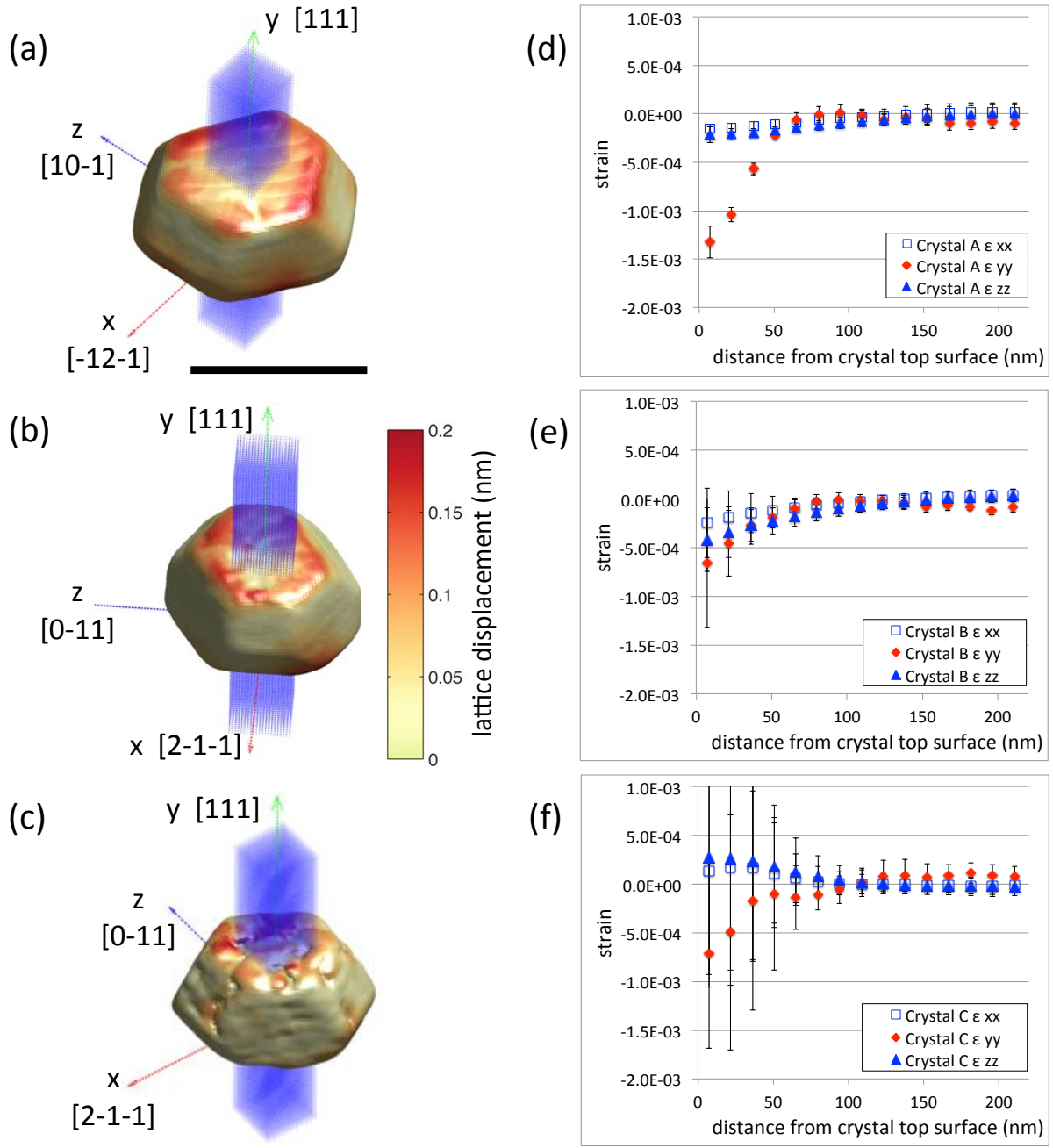

Extended Data Figure 6: Line profiles of normal strain, $\varepsilon_{yy}$, and in-plane strains, $\varepsilon_{xx}$ and $\varepsilon_{zz}$, as a function of distance from the gallium-implanted crystal surface. (a), (b) and (c) 3D morphology of crystals A, B and C respectively coloured by lattice displacement magnitude. Superimposed are points at which strain was extracted for line profile analysis. (d), (e) and (f) Line profiles of $\varepsilon_{xx}$, $\varepsilon_{yy}$ and $\varepsilon_{zz}$ in crystals A, B and C respectively plotted as a function of distance from the implanted surface. The markers show the average strain values at each depth. The error bars capture the standard deviation of $\varepsilon_{xx}$, $\varepsilon_{yy}$ and $\varepsilon_{zz}$ at each specific depth. The scalebar corresponds to 1 $\mu$m.

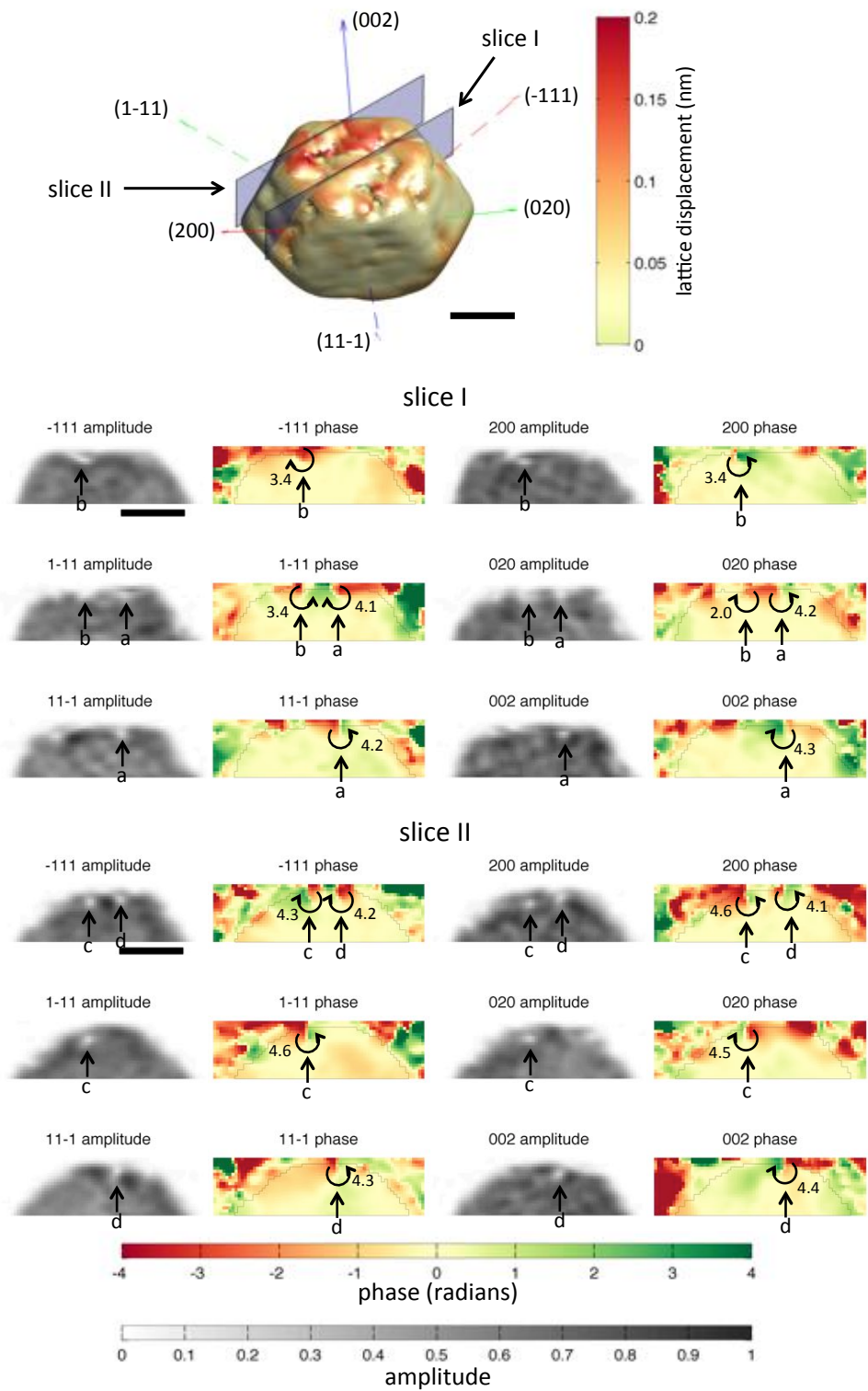

Extended Data Figure 7: 3D rendering of crystal C, coloured according to lattice displacement. Superimposed are the six reflections for which BCDI measurements were carried out. Also shown are two slices for which the amplitude and phase of the complex electron density, recovered from all six lattice reflections, are shown. In both slice I and slice II amplitude and phase features due to defects are visible and four distinct defects have been labelled (a) to (d). The direction of the phase jump associated with specific defects is marked by a circular arrow in the phase maps and its approximate value is noted in radians. A summary of this data is provided in Extended Data Table 2. Scale bars are 300 nm in length.

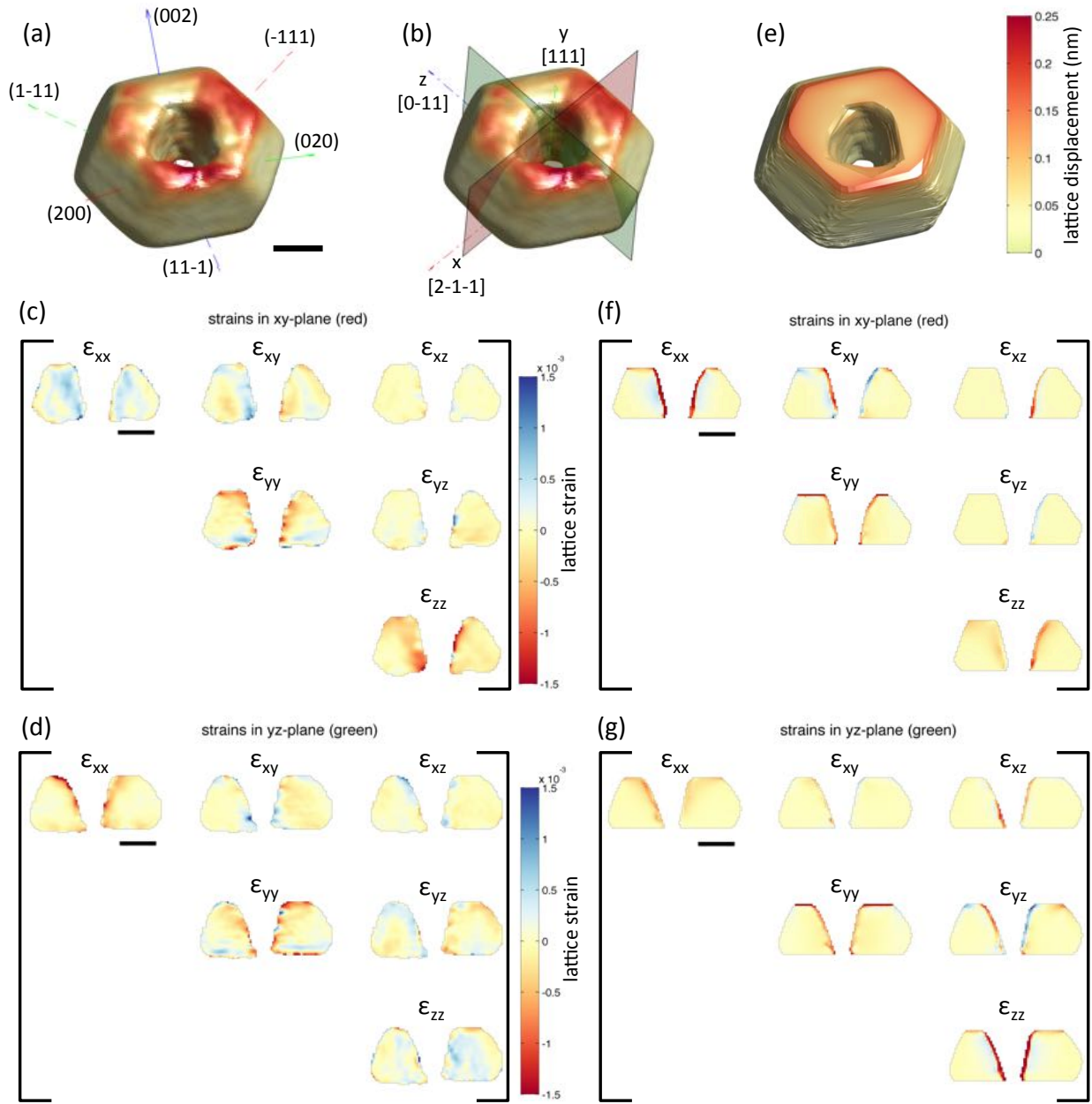

Extended Data Figure 8: Strains induced by extensive FIB milling (crystal D). (a) 3D rendering of crystal D coloured according to the measured lattice displacement magnitude. Superimposed are the six reflections measured using BCDI. (b) Crystal coordinates and sections on which strains are plotted. (c) and (d) Six components of the experimentally measured Cauchy strain tensor plotted on the xy and yz sections respectively that are shown in (b). (e) Finite element model of crystal D coloured according to the calculated lattice displacement magnitude. (f) and (g) Six components of the simulated Cauchy strain tensor field plotted on the same xy and yz sections through the crystal as the experimental results in (c) and (d). Scale bars are 300 nm in length.

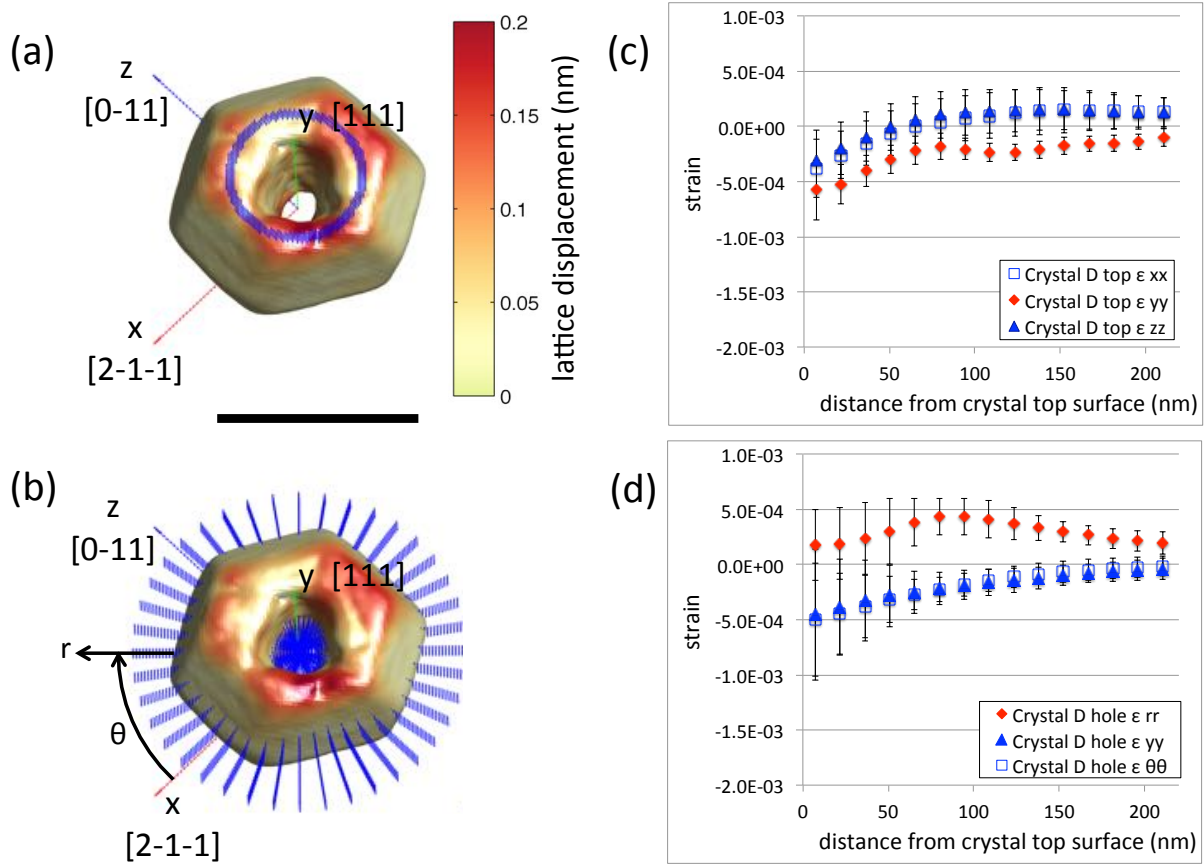

Extended Data Figure 9: Line profiles of strain variation normal to implanted surfaces in crystal D. (a) and (b) show the 3D morphology of crystal D coloured by lattice displacement magnitude. Superimposed are points at which strains were extracted for plotting of the line profiles. The scale bar corresponds to 1 $\mu$m. For the top surface of crystal D, strain was analysed along vertical lines shown in (a). For the milled hole radial lines plotted in (b) were considered. (c) Variation of normal strain, $\varepsilon_{yy}$, and in-plane strains, $\varepsilon_{xx}$ and $\varepsilon_{zz}$, as a function of distance from the implanted top surface of crystal D. The markers show the average strain values at each depth. Error bars capture the standard deviation of $\varepsilon_{xx}$, $\varepsilon_{yy}$ and $\varepsilon_{zz}$ at each depth. (d) Variation of normal strain, $\varepsilon_{rr}$, and in-plane strains, $\varepsilon_{yy}$ and $\varepsilon_{\theta\theta}$, as a function of distance from the implanted surface of the hole milled in crystal D. The markers show the average strain values at each radial distance from the implanted surface. The error bars capture the standard deviation of strains at each radial distance from the implanted surface. The scalebar corresponds to 1 $\mu$m.

11